\RequirePackage{fix-cm}

\documentclass[smallextended]{svjour3}

\usepackage{graphicx}
\usepackage[caption=false]{subfig}
\usepackage{lscape}

\usepackage{natbib}
\usepackage{url}

\usepackage{xspace}
\makeatletter
\xspaceaddexceptions{\check@icr}
\makeatother

\usepackage{wrapfig}

\usepackage{float}

\usepackage{ifpdf}
\ifpdf
       \usepackage[pdftex,colorlinks,pdfauthor={Hugo Buddelmeijer}]{hyperref}
       \DeclareGraphicsExtensions{.pdf,.png,.jpg}
\else
       \usepackage[hypertex,colorlinks]{hyperref}
\fi

\usepackage{algorithm}
\usepackage{algorithmic}

\catcode`~=11 
\newcommand{\urltilde}{\kern -.15em\lower .7ex\hbox{~}\kern .04em}
\catcode`~=13 

\newcommand{\ProcessTargets}{Process Targets\xspace}

\newcommand{\processtarget}{process target\xspace}

\newcommand{\aprocesstarget}{a process target\xspace}
\newcommand{\Aprocesstarget}{A process target\xspace}
\newcommand{\processtargets}{process targets\xspace}

\newcommand{\sourcecollection}{Source Collection\xspace}

\newcommand{\sourcecollections}{Source Collections\xspace}

\newcommand{\SourceCollection}{Source Collection\xspace}
\newcommand{\SourceCollections}{Source Collections\xspace}
\newcommand{\aSourceCollection}{a Source Collection\xspace}

\newcommand{\calculator}{Attribute Calculator\xspace}
\newcommand{\calculators}{Attribute Calculators\xspace}
\newcommand{\Calculators}{Attribute Calculators\xspace}

\newcommand{\AttributeCalculator}{Attribute Calculator\xspace}

\newcommand{\AttributeCalculators}{Attribute Calculators\xspace}

\newcommand{\calculatorDef}{Attribute Calculator Definition\xspace}
\newcommand{\calculatorDefs}{Attribute Calculator Definitions\xspace}

\newcommand{\acalculatorDef}{an Attribute Calculator Definition\xspace}

\newcommand{\python}{Python\xspace}

\newcommand{\AW}{{\sf Astro-WISE}\xspace}

\newcommand{\SourceListWrapper}{SourceList Wrapper\xspace}
\newcommand{\SelectAttributes}{Select Attributes\xspace}
\newcommand{\RenameAttributes}{Rename Attributes\xspace}
\newcommand{\ConcatenateAttributes}{Concatenate Attributes\xspace}

\newcommand{\FilterSources}{Filter Sources\xspace}
\newcommand{\SelectSources}{Select Sour\-ces\xspace}
\newcommand{\ConcatenateSources}{Con\-ca\-te\-na\-te Sour\-ces\xspace}
\newcommand{\CombineSources}{Concatenate Sources\xspace}
\newcommand{\External}{External\xspace}

\newcommand{\RelabelSources}{Relabel Sources\xspace}
\newcommand{\Pass}{Pass\xspace}

\newcommand{\refsec}[1]{section~\ref{#1}\xspace}
\newcommand{\reffig}[1]{Fig.~\ref{#1}\xspace}

\newcommand{\refsecp}[1]{section~\ref{#1}\xspace}
\newcommand{\reffigp}[1]{Fig.~\ref{#1}\xspace}

\newcommand{\citesetrelpap}{Paper II\xspace}

\begin{document}

\title{Automatic Optimized Discovery, Creation and Processing of Astronomical Catalogs}


\author{
Hugo Buddelmeijer
\and Danny Boxhoorn
\and Edwin A. Valentijn
}


\institute{
Hugo Buddelmeijer \at
Kapteyn Astronomical Institute, Postbus 800, 9747 AD, Groningen, The Netherlands \\
\email{buddel@astro.rug.nl}           
           \and
Danny Boxhoorn \at
\email{danny@astro.rug.nl}
           \and
Edwin A. Valentijn \at
\email{valentyn@astro.rug.nl}
}

\date{Received: date / Accepted: date}

\maketitle

\begin{abstract}
We present the design of a novel way of handling astronomical catalogs in \AW in order to achieve the scalability required for the data produced by large scale surveys.
A high level of automation and abstraction is achieved in order to facilitate interoperation with visualization software for interactive exploration.
At the same time flexibility in processing is enhanced and data is shared implicitly between scientists.

This is accomplished by using a data model that primarily stores how catalogs are derived; the contents of the catalogs are only created when necessary and stored only when beneficial for performance.
Discovery of existing catalogs and creation of new catalogs is done through the same process by directly requesting the final set of sources (astronomical objects) and attributes (physical properties) that is required, for example from within visualization software.

New catalogs are automatically created to provide attributes of sources for which no suitable existing catalogs can be found.
These catalogs are defined to contain the new attributes on the largest set of sources the calculation of the attributes is applicable to, facilitating reuse for future data requests.
Subsequently, only those parts of the catalogs that are required for the requested end product are actually processed, ensuring scalability.

The presented mechanisms primarily determine which catalogs are created and what data has to be processed and stored:
the actual processing and storage itself is left to existing functionality of the underlying information system.

\end{abstract}

\section{Introduction}
Billions of astronomical objects are detected in large astronomical surveys, for which thousands of properties are quantified.
The classical way to handle catalog data produced by large surveys, is to create a static relational database with a direct interface to the user.
This classical approach has several conceptual drawbacks.
The catalogs are published in releases as is; there is very limited flexibility in the derivation of the data.
Redoing part of the data reduction entails downloading a large part of the original data and reprocessing it offline.
Scientists require knowledge about the internal representation of the data to access it.

Examples of such an approach are the Sloan Digital Sky Survey (SDSS) \citep{2002cs........2013S,2002cs........2014G} and the WFCAM Science Archive \citep{2008MNRAS.384..637H}.
It is possible to create user-defined tables in the `CasJobs' service\footnote{\url{http://casjobs.sdss.org/CasJobs/}} of SDSS.
These are of a limited size and there is no facility to reprocess the data.
This is an inherently \textit{pushing} or \textit{forward chaining} approach, that is, scientists create derived catalogs in a stepwise fashion---starting from the released catalogs---until they reach their required end product.
The used database queries are stored within the tables, but there is no conceptual information about what the data in the catalogs represent.

This paper discusses the design of novel mechanisms to handle such large catalogs in \AW through request driven processing.
That is, scientists request their required end product, and the information system autonomously determines the best way to provide this catalog.
We achieve a high level of automation and implicit scalability, while enhancing flexibility in processing and sharing of data.
This is done by using an object oriented data model that focuses on storing information about processing; storing the catalog data itself is of secondary importance.

\subsection{\AW}
\label{sec:proctargetintro}
The \AW consortium has designed a new paradigm
and has implemented a fully scalable information system to overcome the huge information avalanche produced by wide-field astronomical surveys \citep{Mwebaze:2009:ATU:1683300.1683752, awpipeline}.
This is achieved by capturing in a generic way the reality of end-to-end survey operations into a conceptual data model which is translated into hierarchical classes.
The model maps all links between dependencies:
objects are stored in the database, which links all data products to their dependencies.
This creates a \textit{dependency graph} with the \textit{full data lineage} of the entire processing chain.

\AW uses the advantages of Object-Oriented Programming (OOP) to process data in the simplest and most powerful ways.
In essence, it turns the objects that represent conventional astronomical science products, into OOP objects, called \textit{\processtargets}.
Every individual science product, such as frame or catalog, is an instantiation of a specific \processtarget class.
Each of these \processtarget instances knows how to process itself to create the data product it represents.
Each \processtarget has associated \textit{processing parameters}, which are configurable parameters that guide the processing of that target.

The most unique aspect of \AW is its ability to process data based on the final desired result to an arbitrary depth.
This data pulling is the heart of \AW and is called \textit{target processing}.
Contrary to the typical case of forward chaining such as in the SDSS CasJobs service, the \AW database links allow the dependency chain to be examined from the intended \textit{\processtarget} all the way back to the raw data.
A target's dependencies are checked to see if it is \textit{up-to-date}: if there is a newer dependency or if the target does not exist, the target is (re)created.

\subsection{A Functional Approach to Catalogs as \ProcessTargets}
\label{sec:introcontributionssc}
Target processing has been incorporated in the image reduction part of the \AW\ information system since its inception \citep{Mwebaze:2009:ATU:1683300.1683752}.
Originally, only a few classes were available in \AW to handle catalog data, of which the SourceList is the most prominent.
The SourceList is primarily used to create catalogs with attributes derived from images and has limited functionality for creating new catalogs from existing catalogs.
In particular such derived catalogs do not have full data lineage and can therefore not be pulled.
Furthermore they can require large amounts of duplication of catalog data, leading to scalability problems.

This paper describes how data lineage and data pulling mechanisms are extended to cover astronomical catalogs with the design of \processtarget classes---which we call \textit{\SourceCollections}---for catalog data.
A \SourceCollection instance represents a collection of sources (astronomical objects) with attributes (or parameters) that quantify physical properties.
There are separate \processtarget classes for different operations to create and manipulate catalogs (\refsec{sec:operatorsummary}).
The \SourceCollection classes take data pulling mechanisms to a higher level than is necessary for images; in particular it is not required to store the catalog data that a \SourceCollection represents in its entirety.

The full data lineage allows any target to be processed at any time for any reason, since the process parameters unambiguously define how to do so.
Ultimately, this means that it is not necessary to process a target completely, or at all.
In a sense, this turns the Object-Oriented approach into a Functional one:
\Aprocesstarget can also be seen as a representation of the operation that is used derive the science product, in addition to seeing it as a representation of the result itself.
The actual processing of the object and storing the result is then optional.
These two viewpoints are equivalent and interchangeable and the contributions in this work stem from this dual perspective:
\begin{enumerate}
\item We allow \SourceCollections to be created---and used as a dependency for other \processtargets---by specifying their data lineage, without requiring them to be processed, unlike other \processtargets in \AW.
\item \begin{sloppypar}Dependency graphs of \SourceCollections are created automatically through data pulling.
These mechanisms create new \SourceCollections in a way that maximizes their reusability for future data pulling requests.\end{sloppypar}
\item We present a novel way to process only the part of a \SourceCollection that is required for the last \processtarget in a dependency graph.
This is done by using the power of backward chaining to temporarily optimize the dependency graph.
\item We use a novel algorithm (\citet{setrelationspaper}, hereafter Paper II) to infer the logical relationships between catalogs from their data lineage directly.
This is required because the exact set of sources that a catalog represents might not be evaluated.
This algorithm is used to find \SourceCollections and for the optimization of dependency graphs.
\item The methods to calculate new attributes from existing attributes are decoupled from their application.
This offers scientists flexibility in implementing their own methods while reinforcing the principles of data pulling.
\item The catalog objects and data pulling mechanisms are designed to be used in query driven visualization \citep{qdvpaper}.
The high level of automation allows the data pulling to be abstracted, which implicitly minimizes the processing required to create the visualized datasets.
\end{enumerate}

\subsection{Outline}
The remainder of the paper is structured as follows.
The \SourceCollection concept is introduced and demonstrated with an example in \refsec{sec:introducingsourcecollections}.
This is followed by a short description of the different \SourceCollection classes that are implemented in \AW in \refsec{sec:ssoperators} and a discussion about storing \SourceCollections and the catalog data they represent in \refsec{sec:sckeystorelineage}.
Subsequently the concept of \textit{dependency graphs} is explained in \refsec{sec:dependencygraphs} and their automatic creation through data pulling in \refsec{sec:scpullingdata}.
The optimization of dependency graphs is discussed in \refsec{sec:sctreemodifications} and their processing in \refsec{sec:scprocessingandstoring}.
A summary and conclusion is provided in \refsec{sec:conclusions}.

\section{Introducing \SourceCollections}
\label{sec:introducingsourcecollections}
A \SourceCollection is an \AW \processtarget (\refsec{sec:proctargetintro}) for the handling of astronomical catalogs.
These catalogs consist of sets of sources and attributes that quantify properties of the sources.
The exact set of sources and the values of the attributes is determined by processing a \SourceCollection.
When we refer to \textit{catalog data}, we mean this processing result.
A \SourceCollection can also be seen as a representation of the action required to derive the catalog data, since a \SourceCollection can be created without being processed.
We refer to this action in a conceptual sense as the \textit{operator} of a \SourceCollection and define separate \processtarget classes for different operations on catalogs (\refsec{sec:operatorsummary}).

Every source in a \SourceCollection has a unique identifier and two \SourceCollections are considered to represent the same sources if and only if the identifiers of their sources are identical.
A source itself can be seen as an object in the computer science reading of the term.
A parametrized property of a source can then be seen as an attribute of such an object.
We will use the term \textit{attribute} instead of \textit{parameter}, which originates from this object oriented approach.
Attributes quantify physical properties of the sources in \aSourceCollection and the set of attributes forms the \SourceCollection dimensions.
The label of an attribute only describes what physical property is represented by the attribute.
This labeling could be standardized, for example with Unified Content Descriptors; for the scope of this paper we will refer to attributes by their name only.

Every \SourceCollection instance is linked to its dependencies, forming a dependency graph all the way to the raw data.
Dependencies of a \SourceCollection that are \SourceCollections themselves are also called its \textit{parents}, because the catalog represented by the \SourceCollection is derived from them.
Such a dependency graph can be visualized as interconnected nodes.
In the figures in this paper, the dependencies of \aprocesstarget are shown above it.
Therefore the data processing runs from top to bottom and the data lineage from bottom to top.
Such a dependency graph can conceptually be extended in both directions. 
The top nodes will contain photometric attributes and can be connected to nodes representing frames that were used to measure these attributes.
The bottom nodes can be connected to nodes representing hypothetical \processtargets for graphs or other analysis products.

\subsection{\SourceCollection Example}
\label{sec:scexample}
\begin{figure}[ht!]
 \centering
 \subfloat[Primitive] 
   {\label{figscintroexampleprimitive}\includegraphics[height=0.16\linewidth]{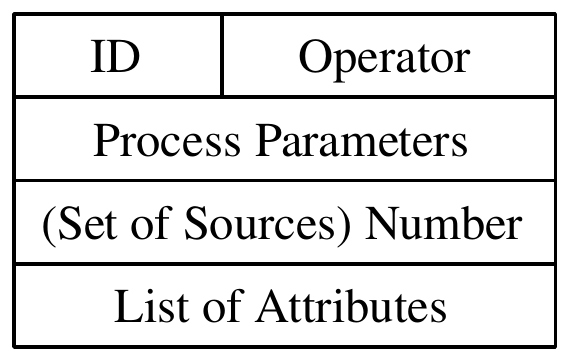}}
 \subfloat[Example]
  {\label{figscintroexamplesc}\includegraphics[height=0.16\linewidth]{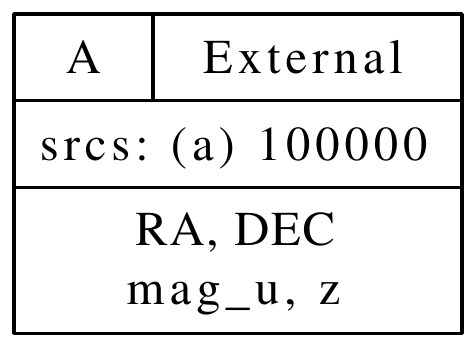}}
\hspace{0.5cm}
 \subfloat[A scientist pulling data from the information system.]
   {\label{figscintroexampleman}\includegraphics[width=0.45\linewidth]{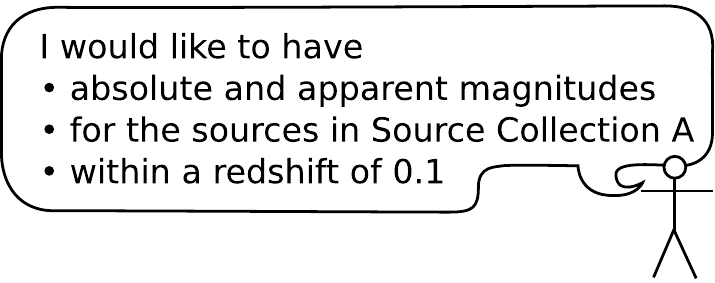}}
 \caption{
(a) A \SourceCollection primitive, (b) a representation of a \SourceCollection used in the example and (c) the scientist formulating a data pulling request.
The following elements can be seen in the \SourceCollection representation:
Top left: a unique identifier of this \SourceCollection.
Top right: the operator of the \SourceCollection.
The second row represents the process parameters, if any.
The next row describes the sources of the \SourceCollection.
The number on the right is the number of sources and the letter between parenthesis represents the exact set of sources.
\SourceCollection with the same symbol represent the exact same set of sources; different symbols might represent different sets.
At the bottom: the names of the attributes that are represented by this \SourceCollection; in this case celestial coordinates, an apparent magnitude in the $u$ band and a redshift.
}
 \label{fig:scintroexample}
\end{figure}

We demonstrate the \SourceCollection concept with a simplified example of data pulling.
We assume the existence of a \SourceCollection (labeled $A$, \reffig{fig:scintroexample}) that contains apparent magnitudes and redshifts for a large set of galaxies.
A scientist pulls a dataset with both absolute and apparent magnitudes for nearby galaxies.
First, the scientist formulates a data pulling request (\reffig{fig:scintroexample}) in which three pieces of information are specified:
\begin{itemize}
\setlength{\itemsep}{0pt}
 \item The data set from which the sources should be selected: \SourceCollection $A$.
 \item The selection criterion for the sources: a redshift below 0.1.
 \item The required attributes: absolute and apparent magnitudes.
\end{itemize}

\begin{figure}[ht!]
 \centering
 \subfloat[Persistently Stored \SourceCollections]{\label{fig:scintreexamplepers}\includegraphics[width=0.45\linewidth]{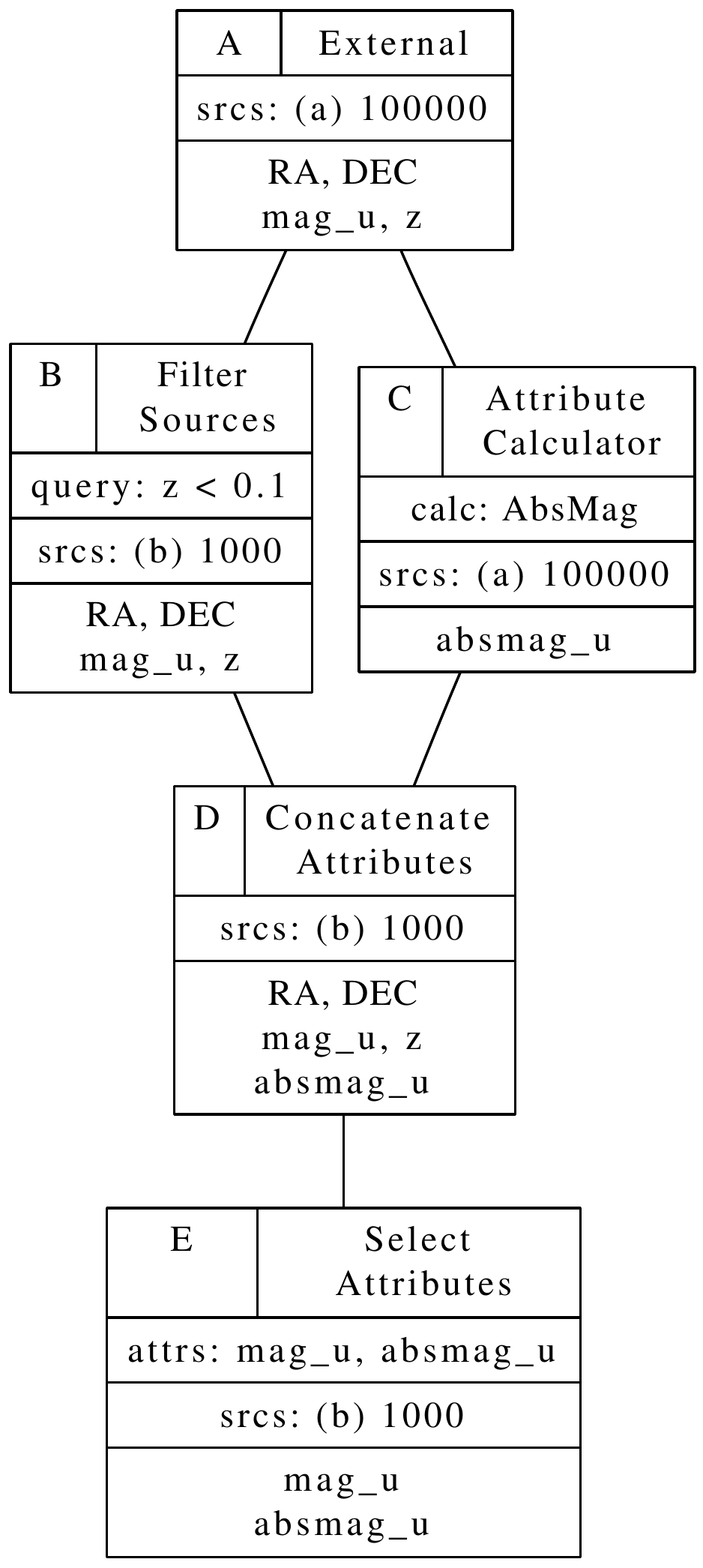}}
 \subfloat[Transient \SourceCollections used for Processing]{\label{fig:scintreexampletrans}\includegraphics[width=0.45\linewidth]{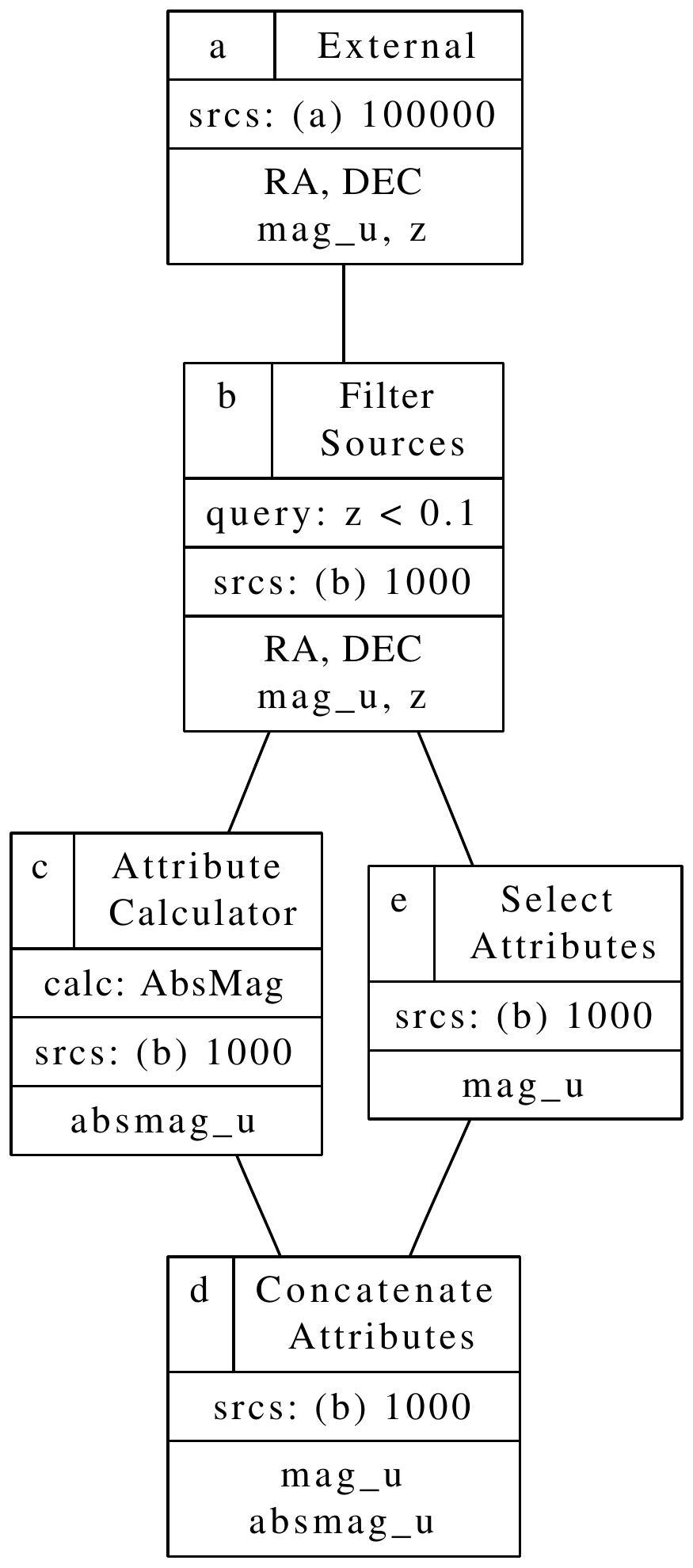}}
 \caption{Two dependency graphs of \SourceCollection, generated by the information system. Every box represents a \SourceCollection.
The \SourceCollections on the left are persistently stored, where the \calculator\ is defined as general as applicable, to facilitate reuse. 
The \SourceCollections on the right are temporary and transient, where the \calculator\ is defined as specific as possible, to minimize the required processing.}
 \label{fig:scintroexample2}
\end{figure}

\noindent Subsequently, the information system creates the required \SourceCollections (\reffig{fig:scintreexamplepers}):
\begin{itemize}
\setlength{\itemsep}{0pt}
 \item \SourceCollection $B$ is created to select all sources that match the given selection criterion.
 \item The information system determines that no absolute magnitudes have been defined for these sources and it creates \SourceCollection $C$ to calculate absolute magnitudes from apparent magnitudes.
\item The information system determines that the calculation can be performed on all sources in \SourceCollection $A$.
Therefore, it optimizes for generality and uses \SourceCollection $A$ as dependency for \SourceCollection $C$, instead of $B$.
The \SourceCollection is not yet processed at this stage.
\item \SourceCollection $D$ and $E$ are created to combine the attributes represented by different \SourceCollections and select the required ones.
\end{itemize}

\noindent Finally, the created dependency graph is optimized and processed (\reffig{fig:scintreexampletrans}): 

\begin{itemize}
\setlength{\itemsep}{0pt}
 \item The information system creates a temporary copy of the dependency graph in order to optimize it for scalability to fulfill the request as quickly as possible.
\item It reorganizes the dependency graph to minimize the required processing by placing the selection of sources before the calculation of absolute magnitudes.
\item The information system retrieves the data of \SourceCollection $b$ and uses this to process \SourceCollection $c$ completely.
The calculated attributes will be stored for future requests as part of \SourceCollection $C$, because they cannot be derived on the fly.
\item The other \SourceCollections are processed on the fly while retrieving the catalog data of \SourceCollection $e$.
The catalog data is subsequently returned to the scientist.
\end{itemize}

\subsection{Key Features in Example}
\label{sec:scfeaturesexample}
The example in section \ref{sec:scexample} highlights the key aspects of the \SourceCollection:
\begin{itemize}
\setlength{\itemsep}{0pt}
 \item Catalog data is pulled and new \SourceCollections are created to compute attributes that do not yet exist (section \ref{sec:scpullingdata}).
 \item The final catalog has full data lineage: any attribute value can be recalculated and the selection criterion is stored (\refsec{sec:sckeystorelineage}).
 \item Calculations are defined to be as general as applicable. \SourceCollection $C$ can be reused if at a later stage absolute magnitudes are requested for another subset of \SourceCollection $A$
(section \ref{sec:scpullingdata}).
 \item The information system reorganizes the order of the \SourceCollections to prevent the calculation of unnecessary data (\refsec{sec:sctreemodifications}).
The algorithm to determine logical relationships between sets of sources of \citesetrelpap is used for more complex dependency graphs.
 \item \SourceCollection $C$ is processed partially by processing its smaller copy $c$ entirely and sharing the result (sections \ref{sec:sckeystorelineage}, \ref{sec:sctreemodifications}).
 \item The calculation of the absolute magnitudes can be performed on the workstation of the scientist or on a distributed computing cluster, while the selection of data can be performed on the database (\refsec{sec:processingscs}).
\end{itemize}

\section{\SourceCollection Classes: Elementary Operations on Catalogs}
\label{sec:operations}
\label{sec:ssoperators}
\label{sec:sckeyoperators}
Many of the novel features of the \SourceCollections originate from the ability of the information system to assess aspects of the catalogs by inspecting only the data lineage.
This is achieved by having a predefined set of operations that can be used to process a \SourceCollection.
Separate \processtarget classes are designed for the different operations.
We use the term \textit{operator} to refer to the action required to create the catalog data.

These operators are designed to be as elementary as possible in order to maximize the information that can be inferred from the data lineage directly.
Therefore, there are no \SourceCollection operators that are entirely user-defined.
However, the behavior of \SourceCollections can be influenced by setting the process parameters.
For example, we do define an operator to calculate new attributes of sources from existing attributes (\refsec{sec:sckeycalculator}).
This allows scientist to specify their own calculation method as a process parameter.

There are two main effects of the elementary operators:
firstly, they allow the information system to determine whether a \sourcecollection can be used in the construction of a dependency graph (section \ref{sec:scpullingdata}).
Secondly, they allow efficient reorganization of the dependency graph, e.g. for partial processing (section \ref{sec:sctreemodifications}).

Most operators we define are modeled after relation operations \citep{Codd:1970:RMD:362384.362685} to allow them to be evaluated on the \AW database. 
In essence, we extend SQL commands to target processing, although this is not directly our goal.
The important aspect in the design of the operators is maximizing the information that can be inferred from the data lineage.
Not all operators we describe can be evaluated on SQL and vice versa, most operators can be evaluated in the \AW \python environment as well.

\subsection{List of classes}
\label{sec:operatorsummary}
We summarize the operators that are most important for our research:
\begin{itemize}
\setlength{\itemsep}{0pt}
 \item \textbf{\SelectAttributes}: Selects a subset of attributes from a parent \SourceCollection.
 \item \textbf{\ConcatenateAttributes}: Combines the different attributes from several parent \SourceCollections that represent the same sources.
 \item \textbf{\RenameAttributes}: Renames attributes of a parent \SourceCollection.
 \item \textbf{\FilterSources}: Selects a subset of sources from a parent \SourceCollection by evaluating a selection criterion.
 \item \textbf{\SelectSources}: Selects a subset of sources from a parent \SourceCollection by listing the required sources explicitly.
 \item \textbf{\ConcatenateSources}: Combines the different sources of several parent \SourceCollections that represent the same attributes.
 \item \textbf{\RelabelSources}: Changes the source identifiers of a parent \SourceCollection.
 \item \textbf{\calculator}: Calculates new attributes from existing attributes for the sources in a parent \SourceCollection (\refsec{sec:sckeycalculator}).
 \item \textbf{\External}: Represents a catalog without data lineage.
 \item \textbf{\Pass}: Represents the exact same catalog as its parent.
 \item \textbf{\SourceListWrapper}: A special \SourceCollection to use the \AW SourceList class as a \SourceCollection. The SourceList class is used to detect sources from images and measure photometric and related attributes.
\end{itemize}

\subsection{Generic Operator for Attribute Calculation}
\label{sec:sckeycalculator}
A special \SourceCollection class is designed for the calculation of new attributes of sources from existing attributes.
The calculation performed by \aSourceCollection of this class, is decoupled from the definition of the class and is stored as another persistent object, which can be created by scientists themselves.

This auxiliary object is called an \textit{\calculatorDef} and contains both information about how to perform the calculation as well as information about the calculation itself: which attributes are calculated, which attributes are required and which process parameters can be set.
This allows the information system to discover attribute derivation methods in order to instantiate \SourceCollections to calculate these attributes for a requested set of sources.
This offers scientists flexibility in implementing their own methods while reinforcing the principles of data pulling.

Multiple \calculatorDefs might exist for the calculation of the same attribute, for example through different methods or different versions of the same method.
\AW has functionality to indicate that stored objects should not be used anymore by invalidating them, for example when a newer version of the object exist.
This is used within the \SourceCollections to indicate that newer versions of \calculatorDefs exist.
This allows existing functionality to be used for ensuring that catalogs are always created with the latest method and that out-dated catalogs are flagged for possible recreation.

\section{Storing Data Lineage instead of Tables}
\label{sec:sckeystorelineage}
SourceCollections can be created and stored by specifying their data lineage only; it is not required to process them.
That is, the actual determination of the exact composition of sources in a catalog, and the calculation of the values of their attributes, is delayed as long as possible.
Furthermore, the result of the processing is stored only if necessary for performance reasons and the results can be shared between \SourceCollections.
We summarize the benefits of this approach:
\begin{itemize}
\item Different \SourceCollection can represent partially identical catalogs without any duplication of stored data.
\item The processing of intermediate \SourceCollection can be limited to those subsets that are required for the end node of a dependency graph.
\SourceCollections can therefore be created with arbitrary sizes without performance penalties.
This ensures maximum reusability of the created \SourceCollections.
\item No results have to be stored at all for \SourceCollections that can be processed on the fly.
\end{itemize}

\subsection{\SourceCollection Persistent Properties}
\label{sec:definitionofsc}
The \textit{persistent properties} of \aprocesstarget are the properties of the object that are stored in a database.
These properties can be grouped in the following types, a categorization that is especially important for \SourceCollections:
\begin{itemize}
\setlength{\itemsep}{0pt}
 \item \textbf{Data Lineage}: Properties that define the catalog that is represented by the \SourceCollection.
These are dependencies and process parameters.
Dependencies are other \processtargets from which the catalog represented by this \SourceCollection is derived, often \SourceCollections as well.
Process parameters that influence the processing as defined by the class of the \SourceCollection.
The dependencies and process parameters together unambiguously define the catalog that the \SourceCollection represents.
\item \textbf{Processing Results}: Results of processing the \SourceCollection, detailed in section \ref{sec:processingresults}.
 \item \textbf{Other Properties}: Properties that do not refer directly to the processing or the processing results.
These include identifiers of the object, a human readable name of the \SourceCollection, a reference to its creator, status of the processing, etc. Some of these can be specified by the user, others are set automatically by the information system.
\end{itemize}

\subsection{Processing Results}
\label{sec:processingresults}
The result of processing \aprocesstarget instance (\refsec{sec:proctargetintro}) can be stored persistently.
The processing results of image classes are primarily the values of the pixels of the image, which in \AW are stored as FITS files on the dataserver.
For \SourceCollections the primary result is the catalog data it represents, which in \AW is stored in the database.

The \SourceCollection classes are designed to allow partial processing of objects, for example because only a part of the catalog data is required at a specific moment.
The processing results are split up in distinct components in order to achieve this.
These components can, in principle, be processed separately.
The following results can be distinguished:
\begin{itemize}
 \item The catalog the \SourceCollection represents: the values of all the attributes for all the sources. This is the primary processing result and can be decomposed in the partial results that follow. 
 \item A partial catalog: the values of the attributes for a subset of the sources or attributes.
 \item The set of sources the \SourceCollection represents, which can be seen as a list of identifiers of the sources. This can be further split up into the number of sources, or an identification of the set without actually enumerating all the sources individually.
 \item The set of attributes of the sources. That is, which physical properties the \SourceCollection represents, not the actual values of the attributes.
\end{itemize}
To process a \SourceCollection partially, a new \processtarget is created that only represents the required component, which is subsequently processed in its entirety.
Such a component is either stored in its entirety or not at all, and can be shared between \SourceCollections.

The sharing of processing results leads to multiple paths to the same stored data.
The dependency graphs representing these different paths are only created automatically by the information system through modifications of existing dependency graphs (\refsec{sec:graphmodification}).
The information system ensures that the different paths are equivalent by only performing modifications where this is guaranteed.

\section{\SourceCollection Dependency Graphs}
\label{sec:dependencygraphs}
A \SourceCollection represents a catalog that is derived from its dependencies, which again have dependencies themselves.
These dependencies chain a \SourceCollection back to the raw data and form a graph of \processtargets.
The term \textit{dependency graph} is used to refer to this complete set of dependencies of a \SourceCollection.
These graphs are \textit{directed acyclic graphs}, or \textit{acyclic digraphs}, because there are no cyclic dependencies \citep{Thulasiraman92}.

In the figures depicting dependency graphs in this paper, the dependencies of a \SourceCollection are shown above it.
Therefore the data processing runs from top to bottom and the data lineage from bottom to top.
There are no arrows on the shown edges, because the preferred direction is dependent on context.
This paper only treats the part of such a dependency graph that considers \SourceCollections.

\subsection{Modifications of Dependency Graphs}
\label{sec:graphmodification}
The information system can modify dependency graphs of \SourceCollections, e.g. while constructing new ones or when optimizing existing ones as discussed in the next sections.
All modification steps in the following algorithms are performed by replacing a \SourceCollection with another one.
There are two ways to do this:
\begin{itemize}
 \item Replacing a \SourceCollection with another one that represents the exact same catalog. This is the only mechanism that is used in the dependency graph optimization (section \ref{sec:sctreemodifications}).
 \item Replacing a \SourceCollection with one that represents a different catalog. This is only performed during the creation of new dependency graphs (section \ref{sec:sctreemodifications}) and only on dependencies of the \Pass \SourceCollections at the end of the graph.
\end{itemize}

The individual modifications themselves are designed in a way that separates the knowledge of \textit{how} to perform a modification and \textit{why} to do so.
How to perform a modification is part of the definition of the \SourceCollection classes.
Whether a specific modification should be applied is the responsibility of the part of the information system that governs the entire dependency graph.
Therefore, all modifications are between a \sourcecollection and its direct dependencies, because an individual \sourcecollection has no knowledge of other objects.

A specific kind of modifying a dependency graph is `moving' \SourceCollections through the graph.
The way this should be interpreted---in simplified form---is that copies of a \SourceCollection and its parent are created, but with their dependencies swapped.
The original \SourceCollection is then replaced by these copies.
As a result, \sourcecollections can only be moved `up' the graph.
To move a \sourcecollection down, the \sourcecollection with that \sourcecollection as a parent should be moved up.

Some modifications can only be performed if the relationship between the sets of sources of the involved \sourcecollections is known.
The information system uses the algorithm of \citesetrelpap to provide this information to the individual \sourcecollections.

\section{Pulling Catalogs}
\label{sec:scpullingdata}
The `pushing' way to use catalogs such as represented by the \SourceCollections is to define the catalog, process and store it, and then request subsets of the catalog.
This order is changed with target processing \citep{Mwebaze:2009:ATU:1683300.1683752}.

\SourceCollections are primarily created automatically by pulling data, which means that the evaluation of processing starts at the end of the chain by requesting the final catalog that is required.
The information system will autonomously create a dependency graph of \SourceCollection which ends with a \SourceCollection that represents the requested catalog (Algorithm \ref{algo:pullcatalogtree}).

There are two main goals of the data pulling mechanisms with respect to the creation of the dependency graph:
Firstly, they ensure that existing \SourceCollections will be reused as much as possible and secondly, new \SourceCollections are created in a way that maximizes their reusability.

\subsection{Data Pulling: Formulating a Request}
The pulling of data starts with a request for a specific dataset.
In our research we have limited such requests to three pieces of information:
\begin{itemize}
\setlength{\itemsep}{0pt}
 \item A starting \SourceCollection from which a selection is made.
 \item A list of required attributes, not necessarily represented by the starting \SourceCollection.
 \item Optionally, a selection criterion for the sources.
\end{itemize}

\begin{algorithm}[hbtp]
\caption{Creating Target Dependency Graph}
\label{algo:pullcatalogtree}
\begin{algorithmic}[1]
\STATE Receive and parse a request for catalog data.
\STATE Instantiate the starting \SourceCollection.
\STATE Create a \SelectAttributes that selects an empty attribute list from this \SourceCollection.
\STATE Create a \Pass \SourceCollection with the \SelectAttributes as parent.
\IF{a selection criterion is specified}
\STATE Select the right sources (Algorithm \ref{algo:pullcatalogselect}).
\ENDIF
\FORALL{requested attributes}
\STATE Add the attribute to the \Pass \SourceCollection (Algorithm \ref{algo:pullcatalogaddattribute}).
\ENDFOR
\end{algorithmic}
\end{algorithm}

\subsection[Derivation Preferences]{Data Pulling: Derivation Preferences}
The information system will use existing and newly created \SourceCollections to create a dependency graph which ends with a \SourceCollection representing the requested catalog.
The information system is able to autonomously decide how to proceed if there are multiple \SourceCollections that can be used to fulfill a particular dependency.
This is done by applying a ranking function to all \SourceCollections that can be used and select the one with the highest ranking.
Scientists can influence this process by specifying their own ranking function or by overruling the choices made by the information system manually. 

\subsection[Selecting Sources]{Data Pulling: Selecting Sources}
Fulfilling a request for a catalog begins with creating a \SourceCollection with the correct the composition of sources (Algorithm \ref{algo:pullcatalogselect}).
The resulting \SourceCollection will only represent the selected set of sources at this stage, without attributes. 

In this paper we restrict ourselves to requesting subsets of sources that are already represented by an existing \SourceCollection, because our focus is on operations on catalogs.
In particular we assume the existence of \SourceCollections with photometric and related attributes derived from images.
These catalogs could be created through pulling mechanisms as well; this is beyond the scope of this paper.

The logical relations algorithm of \citesetrelpap is used to search for an existing \SourceCollection that represents the requested selection.
First all \SourceCollections that represent the same sources as the original \SourceCollection are found.
Subsequently a \FilterSources is sought, one with the specified selection criterion and with one of these \SourceCollections as parent.
New \SourceCollections are created to select the required sources if no suitable \SourceCollection is found.
This might require more than only a single \FilterSources \SourceCollection because the information system has to ensure that the attributes used in the selection criteria are available.

For example, the specified selection criterion in the example in \refsecp{sec:scexample} depends on the availability of the redshift attribute.
The information system would have tried to find this attribute if it would not have been included in \SourceCollection $A$.

A \SelectAttributes \SourceCollection is created to select no attributes from the found or created \SourceCollection with the sources.
The required attributes are subsequently added to this new \SourceCollection with only sources (\refsec{sec:selectingattributes}).

\begin{algorithm}[hbtp]
\caption{Selecting Sources}
\label{algo:pullcatalogselect}
\begin{algorithmic}[1]
\STATE Search for all \SourceCollections representing the original sources.
\STATE Search for all \FilterSources with one of these \SourceCollections as parent and the specified criterion as parameter.
\STATE Rank all found \SourceCollections.
\IF{a suitable \SourceCollection is found}
\STATE Use the highest ranking \SourceCollection to represent the sources.
\ELSE
\STATE Use algorithm \ref{algo:pullcatalogtree} to create a \SourceCollection with all attributes referenced in the selection criterion.
\STATE Create a new \FilterSources to represent the sources.
\ENDIF
\STATE Create a \SelectAttributes to select no attributes from the \SourceCollection representing the sources.
\STATE Create a \SelectSources with the original \SourceCollection as parent and the \SelectAttributes to specify the selected sources.
\STATE Use the \SelectSources as the parent of the final \Pass \SourceCollection.
\end{algorithmic}
\end{algorithm}

\subsection[Selection Attributes]{Data Pulling: Selecting Attributes}
\label{sec:selectingattributes}
A catalog pulling request should contain a list of required attributes.
For every requested attribute, the information system will search for an existing \SourceCollection that represents this attribute for the requested sources.
A hierarchy of \SelectAttributes and \ConcatenateAttributes \SourceCollections is created to add the attribute to the \Pass \SourceCollection already representing the sources (Algorithm \ref{algo:pullcatalogaddattribute}).

Requested attributes for which no suitable \SourceCollections can be found, are derived with new \SourceCollections (Algorithm \ref{algo:pullcatalognewcalculators}).
In this paper we limit ourselves to attributes that are derived from other attributes using \calculators \SourceCollections.
The calculation performed by an \AttributeCalculator is specified through a process parameter referencing \acalculatorDef object.
New \AttributeCalculator \SourceCollections are instantiated for all \calculatorDefs that can be used to derive the requested attribute.
The search for attributes is applied recursively if more attributes are required for the derivation of the requested attributes, as specified by the \calculatorDef.

\SourceCollections that require the calculation of new attributes will automatically be defined to operate on the largest dataset the calculation is applicable for.
This is done by giving \SourceCollections which represent a larger set of sources a higher ranking when searching for attributes.

\begin{algorithm}[hbtp]
\caption{Adding Attributes}
\label{algo:pullcatalogaddattribute}
\begin{algorithmic}[1]
\STATE Search for all \SourceCollections representing the attribute.
\STATE Rank all found \SourceCollections.
\IF{a suitable \SourceCollection is found}
\STATE Use the highest ranking \SourceCollection to represent the attribute.
\ELSE
\STATE Create an \AttributeCalculator to represent the attribute (Algorithm \ref{algo:pullcatalognewcalculators}).
\ENDIF
\STATE Create a \SelectAttributes that selects the requested attribute from this \SourceCollection.
\STATE Create a \ConcatenateAttributes with the original parent of the final \Pass \SourceCollection and the new \SelectAttributes as parents.
\STATE Use the \ConcatenateAttributes as new parent of the final \Pass.
\end{algorithmic}
\end{algorithm}

\begin{algorithm}[hbtp]
\caption{Instantiate \Calculators}
\label{algo:pullcatalognewcalculators}
\begin{algorithmic}[1]
\STATE Search for all \calculatorDefs that can be used to calculate the required attribute.
\FORALL{found \calculatorDefs}
\STATE Create a \SourceCollection with all the attributes required by the \calculatorDef (Algorithm \ref{algo:pullcatalogtree}).
\STATE Create an \AttributeCalculator with that \SourceCollection as parent, using the \calculatorDef.
\ENDFOR
\STATE Rank all created \AttributeCalculators.
\STATE Use the highest ranking \AttributeCalculator to represent the attribute.
\end{algorithmic}
\end{algorithm}

\subsection[Catalog Processing]{Data Pulling: Storing \SourceCollections}
The result of data pulling is the creation of a dependency graph that ends with a \SourceCollection that represents the requested catalog.
The \SourceCollections in this dependency graph might be stored persistently if necessary.
The information system will subsequently optimize this graph to process it in the most optimal way (\refsec{sec:modifyingsc}).

\section{Optimization of Dependency Graphs and Processing}
\label{sec:processingscs}
\label{sec:modifyingsc}
\label{sec:sctreemodifications}
\label{sec:sclazydata}
\label{sec:sccachingdata}
The information system will optimize the dependency graph of a \SourceCollection before processing the \SourceCollections that it contains.
There are two goals to these optimizations: minimization of the required processing and optimization of the processing itself.
These optimizations are performed on a temporary transient copy of the dependency graph, which can be discarded after the processing is completed.

Reducing the necessary processing to the minimum required for the last \SourceCollection is the primary goal of this paper.
In essence this is done by placing filtering \SourceCollections before \SourceCollections that create new attributes and removing parts of the dependency graph that are not necessary for the final result.
This will ensure that the \SourceCollections in the dependency graph only represent data that is required for the requested catalog data.
These mechanisms allow the information system, or the scientist, to create and store \SourceCollections instances in their most general and reusable form, (e.g. as in \reffig{fig:scintreexamplepers}), because the creation and storage of the catalog data is minimized automatically (e.g. as in \reffig{fig:scintreexampletrans}).

Optimization of the processing itself is a secondary goal of this paper.
This is done by reorganizing the dependency graphs such that the processing can be performed on the most suitable subsystem of the information system.
For example, \SourceCollections that can best be processed on the database are placed such that they can be combined into one SQL query and processed together.
Parts of the dependency graph can be parallelized in order to process large \SourceCollections on a distributed cluster, especially those that cannot be processed on the database.

Optimizing the dependency graph of the example in figure \ref{fig:scintroexample2} is depicted in figure \ref{fig:scintreexampledata}.
The required processing in this example is dominated by a calculation of absolute magnitudes.
Without optimization, absolute magnitudes have to be calculated for 100\thinspace000 sources; with optimization the calculation is only performed for the 1000 sources that are actually requested, resulting in a factor 100 increase in performance.
The optimizations required to determine the exact set of sources in a \SourceCollection is depicted in figure \ref{fig:scintreexamplesources}.
In this case, the calculation of absolute magnitudes is removed from the dependency graph entirely and the entire graph can be processed on the database.

\subsection{Dependency Optimization: Strategy}
\label{sec:optimizationstrategy}
The best strategy for the optimization of dependency graphs depends on many factors, such as the size of the catalogs, how they will be processed, etc.
Therefore it is not possible to give a one-size-fits-all optimization strategy.
Algorithms \ref{algo:optimizeforload} and \ref{algo:simplifyforload} are procedures that cover most scenarios, they can be adjusted for particular cases.
The steps described in the algorithms are detailed in the rest of the section.

\begin{algorithm}[hbtp]
\caption{Optimization for Processing}
\label{algo:optimizeforload}
\begin{algorithmic}[1]
\STATE Create transient copy of the involved \SourceCollections.
\STATE Simplify the dependency graph (Algorithm \ref{algo:simplifyforload}). Perform this step after every movement-step.
\STATE Move all \SelectAttributes up the graph, to remove parts of the graph.
\STATE Convert all \FilterSources to \SelectSources, to move them through the graph.
\STATE Move all \SelectSources down, to copy them to every part of the graph.
\STATE Move all \SelectSources up the graph, to limit processing to the required subset.
\STATE Move all \SelectAttributes up the graph once more, to simplify the graph further.
\STATE Move all \SelectSources up the graph in order to combine them.
\end{algorithmic}
\end{algorithm}

\begin{algorithm}[hbtp]
\caption{Simplification for Processing}
\label{algo:simplifyforload}
\begin{algorithmic}[1]
\REPEAT
\STATE Convert processed \SourceCollections to \External \SourceCollections.
\STATE Remove parts of the graph with unnecessary dependencies.
\STATE Remove \SourceCollections that are essentially a \Pass \SourceCollection.
\STATE Integrate \SourceCollections and their parents if possible.
\STATE Unite identical \SourceCollections, especially those with the same parents.
\UNTIL no more of these modifications are possible.
\end{algorithmic}
\end{algorithm}

\subsection{Dependency Optimization: Simplifications}
A dependency graph of \SourceCollections can be simplified as part of the optimization routines (Algorithm \ref{algo:simplifyforload}).
A \SourceCollection that has already been fully processed does not have to be processed again.
All these \SourceCollections can be substituted with an \External \sourcecollection that represents the same catalog.
Furthermore, the complexity of a dependency graph can be reduced by combining operators or removing redundant ones.

For example, the initial \SourceCollection in \reffigp{fig:scintroexample2} is an \External \SourceCollection for simplicity.
In a realistic scenario, this \SourceCollection would have dependencies of its own and would only be substituted with an \External \SourceCollection just before processing.
In \reffigp{fig:scintreexamplesources1} two serial \SelectAttributes \SourceCollections are combined and in \reffigp{fig:scintreexampledata4} two parallel \SelectSources \SourceCollections are combined.

\subsection{Dependency Optimization: Removing Dependencies}
Unnecessary parts of a dependency graph can be removed by moving \SelectAttributes \SourceCollections up in the graph (Algorithm \ref{algo:optimizeforload}).
The result of moving a \SelectAttributes up past a \ConcatenateAttributes, might be that one of the dependencies of the \ConcatenateAttributes does not represent attributes anymore.
The part of the graph that ends with this dependency might be removed from the graph in its entirety.

The set of sources of a \ConcatenateAttributes is the intersection of the sets of sources of its parents.
Therefore it is only possible to remove this part of the dependency graph if doing so does not influence the selection of sources.
The logical relations algorithm of \citesetrelpap is used to determine whether this is the case.

\subsection{Dependency Optimization: Sources Limitation}
\label{sec:optimization:sourceslimitation}
\SelectSources \SourceCollections are moved through the dependency graph to ensure that 
only those parts of the \SourceCollections are processed that are required to create the catalog data of the end node (Algorithm \ref{algo:optimizeforload}).
\FilterSources \SourceCollections first have to be converted into a \SelectSources.
Before moving the \SelectSources \SourceCollections up the dependency graph, they are moved down in order to copy them to all parts of the dependency graph they are applicable to.

In \reffigp{fig:scintreexampledata} the \SelectSources \SourceCollection is moved up through the \calculator \SourceCollection.
This creates a copy of the \calculator that represents a subset of the sources of the original.

\subsection{Dependency Optimization: Parallelization}
\label{sec:treeparallelization}
The \SourceCollections are well suited for parallelization because they are processed on a per-source basis.
A \SourceCollection can be parallelized by creating a set of \SelectSources (or \FilterSources) \SourceCollections that each select a subset of the original target, such that all sources are selected exactly once.
The set of \SelectSources \SourceCollections is then combined with a \CombineSources \SourceCollection which can replace the original \SourceCollection in the dependency graph.
Further optimization can move the \SelectSources up to parallelize the entire graph.
The parallelization algorithm is currently not implemented in \AW.

\section{Processing and Storage}
\label{sec:scprocessingandstoring}
The result of the dependency graph optimization is a set of \SourceCollections that requires the least amount of processing to create the requested catalog data.
The information system will recursively process the \SourceCollections and store the results if necessary.

The mechanisms designed for the research presented in this paper are intended be used in conjunction with existing large-scale data storage and processing facilities.
Therefore, the precise way catalog data is processed and stored is largely beyond the scope of this paper and will depend on what is available in the information system.
We give a general discussion of how the processing and storage could be achieved and highlight how this is implemented in \AW.

In particular, for this paper we assume the existence of mechanisms for authentication of users, privilege management and for queuing requests when processing \SourceCollections on shared resources.

\subsection{Processing: Processing \SourceCollections}
\label{sec:derivingandstoringdata}
\label{sec:scviewsandbackends}
The information system can process the \SourceCollections on the most suitable subsystem to achieve scalability for large scale catalogs and real time interaction for small scale catalogs.
We describe the different subsystems to evaluate the operators on:
\begin{itemize}
\item \textbf{Database}:
The selection and combining operators are designed to be evaluated on a database.
The operators of consecutive \SourceCollections can be combined into one database query.
The database can create indexes on columns containing attributes that are frequently used in selection criteria and can automatically cache the results.
Some \SourceCollections, in particular \AttributeCalculators, will not be suitable to be processed on the database.

\item \textbf{Workstation}:
The processing can be performed on the workstation of the scientist for \SourceCollection that cannot be processed on the database.
Furthermore, all the relational operators should also be evaluated on the local machine during interactive visualization of small datasets.
The latency of a round trip to the database or distributed computing facility is too large for responsive interaction.
Such a local implementation of the \SourceCollections holds all the catalog data in memory or in files.

\item \textbf{Distributed Computation}:
Operators that require large computations can be performed on a distributed processing cluster.
This is done by parallelizing the respective parts of the dependency graph and evaluating each sub-graph on a cluster node.

\end{itemize}
Within \AW, most operators can be performed on both the Oracle 11g database or in the Python.
Database queries usually scale linearly; requests similar to the example of section \ref{sec:scexample} are typically delivered with speeds of 100\thinspace000 source attributes per second in the current setup.
There is currently no explicit functionality to process \SourceCollections on the distributed processing cluster.

\subsection[Storing Catalog]{Processing: Storing Catalog Data}
\label{sec:subsubstoringdata}
The result of processing a \SourceCollection---the exact set of sources and the values of the attributes---only has to be stored if this is necessary for performance reasons.
Therefore we make no explicit distinction between storing and caching of catalog data.

The optimization process (\refsec{sec:sctreemodifications}) creates copies of \SourceCollections that represent subsets of the originals.
If the information system decides to store the processing result of such a copy, it will append the catalog data of the copy to that of the original.
Deciding what should be stored is primarily the responsibility of the information system.
The decision should be made for individual processing results.

\label{sec:sccachingsources}
For example, it can be useful to store the identifiers of the sources without storing the values of the attributes in order to store the result of evaluating a complex selection criterion.
Different \SourceCollections that represent the same sources can share this processing result.

A key principle of the presented research, inherited from \AW, is that \aSourceCollection cannot be altered once stored.
Therefore, any stored catalog data of a \SourceCollection cannot change either.
Reliability of the data storage, e.g. ACID properties \citep{Gray1981}, is automatically achieved as a result.
For example, an incomplete database transaction will not leave the database in an invalid state because these will only append data that could also be ingested partially in the first place.

Stored catalog data can, in principle, also be deleted at any time, since it can always be recreated.
A deletion mechanism is currently not incorporated in \AW; instead, all catalog data is backed up regularly to be able to recover quickly from database failures.

\subsection[Retrieving Catalog Data]{Processing: Retrieving Catalog Data}
The last node in the dependency graph represents the requested catalog.
Once it has been processed, the catalog data can be returned to the scientist or used for further analysis or visualization.
Any temporary transient \SourceCollections instantiated for the processing are discarded.

\section{Summary and Conclusions}
\label{sec:conclusions}
The presented work shows a novel approach for the handling of source catalogs, as incorporated in \AW.
The core difference between the \AW approach and the way astronomical catalogs are traditionally disclosed, is that the user works with data models rather than a set of tables in a relation database. 
We showed how data pulling is extended to source catalogs, a first step to data pulling and data lineage in the analysis domain.
\Aprocesstarget---labeled a \SourceCollection---is designed to represent catalog data and operations thereon. 
We summarize the key features of our design:
\begin{itemize}
 \item \SourceCollections are primarily created automatically by the information system through data pulling.
 \SourceCollections that derive new data are created as general as possible in order to facilitate reuse and to prevent duplication of data.
 \item \SourceCollections allow a functional approach to target processing: they can be seen as the operation to create the catalog data.
A \SourceCollection is only processed when this is required, not necessarily at the moment it is created.
Every \SourceCollection class correspond to an elementary operation on catalogs; complex operations should be split over multiple \SourceCollection instances.
 \item The \SourceCollection have full data lineage, which allows the information system to assess aspects of the catalogs without processing them. 
For example, it allows the information system to optimize the a dependency graph of \SourceCollections before processing it.
 \item A \SourceCollection is processed by creating temporary copy of the dependency graph and reordering the dependencies so they are as specific as possible in order to minimize the required processing.
 \item A generic \SourceCollection class is designed for the calculation of new attributes from existing attributes. This offers a framework for scientists to implement their own methods while enforcing the benefits of full data lineage and data pulling.
\end{itemize}
The \AW way of handling astronomical catalogs takes care of most of the administrative tasks automatically.
Discovery of existing catalogs and creation of new catalog is done in the same way, by requesting the required end product.
Catalogs are shared implicitly, because existing catalogs are discovered automatically.
New catalogs are automatically created in their most general form, but only the necessary parts are processed.
Together, this allows scientists to focus on the data itself and the science they want to perform instead of how the data is handled.

\begin{acknowledgements}
This research is part of the project ``Astrovis'', research program STARE
(STAR E-Science), funded by the Dutch National Science Foundation (NWO),
project no.\ 643.200.501.
\end{acknowledgements}


\begin{thebibliography}{}
\bibitem[Buddelmeijer et al.(2011a)]{setrelationspaper}Buddelmeijer, H., Valentijn, E.A.: Leveraging data lineage to infer logical relations between sets. in prep. (2011a) (Paper II)
\bibitem[Buddelmeijer et al.(2011b)]{qdvpaper}Buddelmeijer, H., Valentijn, E.A.: Query Driven Visualization of Astronomical Catalogs. {\tt arXiv:1110.2294}
\bibitem[Codd(1970)]{Codd:1970:RMD:362384.362685}Codd, E.F.: A relational model of data for large shared data banks. Commun. ACM 13, 377–387 (1970)
\bibitem[Gray(1981)]{Gray1981}Gray, J.: The Transaction Concept: Virtues and Limitations. Proceedings of the 7th International Conference on Very Large Databases. 144–154 (1981) 
\bibitem[Gray et al.(2002)]{2002cs........2014G}Gray, J., Szalay, A.S., Thakar, A.R., Kunszt, P.Z., Stoughton, C., Slutz, D., van den Berg, J.: Data Mining the SDSS SkyServer Database. {\tt arXiv:cs/0202014}
\bibitem[Hambly et al.(2008)]{2008MNRAS.384..637H}Hambly, N.C., Collins, R.S., Cross, N.J.G., Mann, R.G., Read, M.A., Sutorius, E.T.W., Bond, I., Bryant, J., Emerson, J.P., Lawrence, A., Rimoldini, L., Stewart, J.M., Williams, P.M., Adamson, A., Hirst, P., Dye, S., Warren, S.J.: The WFCAM Science Archive. MNRAS384, 637–662 (2008)
\bibitem[McFarland et al.(2010)]{awpipeline}McFarland, J.: Astro-WISE: an Information System for Wide-field Imaging Surveys. in prep. (2010)
\bibitem[Mwebaze et al.(2009)]{Mwebaze:2009:ATU:1683300.1683752}Mwebaze, J., Boxhoorn, D., Valentijn, E.A.: Astro-wise: Tracing and using lineage for scientific data processing. In: Proceedings of the 2009 International Conference on Network-Based Information Systems, NBIS ’09, pp. 475–480. IEEE Computer Society, Washington, DC, USA (2009)
\bibitem[Szalay et al.(2002)]{2002cs........2013S}Szalay, A.S., Gray, J., Thakar, A.R., Kunszt, P.Z., Malik, T., Raddick, J., Stoughton, C., vandenBerg, J.: The SDSS SkyServer: Public Access to the Sloan Digital Sky Server Data. {\tt arXiv:cs/0202013}
\bibitem[Thulasiraman et al.(1992)]{Thulasiraman92}Thulasiraman, K., Swamy, M.N.S.: Graphs: Theory and Algorithms. Wiley-Interscience (1992)
\end{thebibliography}

\begin{landscape}

\newcommand{\widthscexampleDa}{0.18\linewidth}
\newcommand{\widthscexampleDb}{0.13\linewidth}
\newcommand{\widthscexampleDc}{0.17\linewidth}
\newcommand{\widthscexampleDd}{0.17\linewidth}
\newcommand{\widthscexampleDe}{0.12\linewidth}

\begin{figure}[ht!]
\centering
 \subfloat[]{\label{fig:scintreexampledata1}\includegraphics[width=\widthscexampleDa]{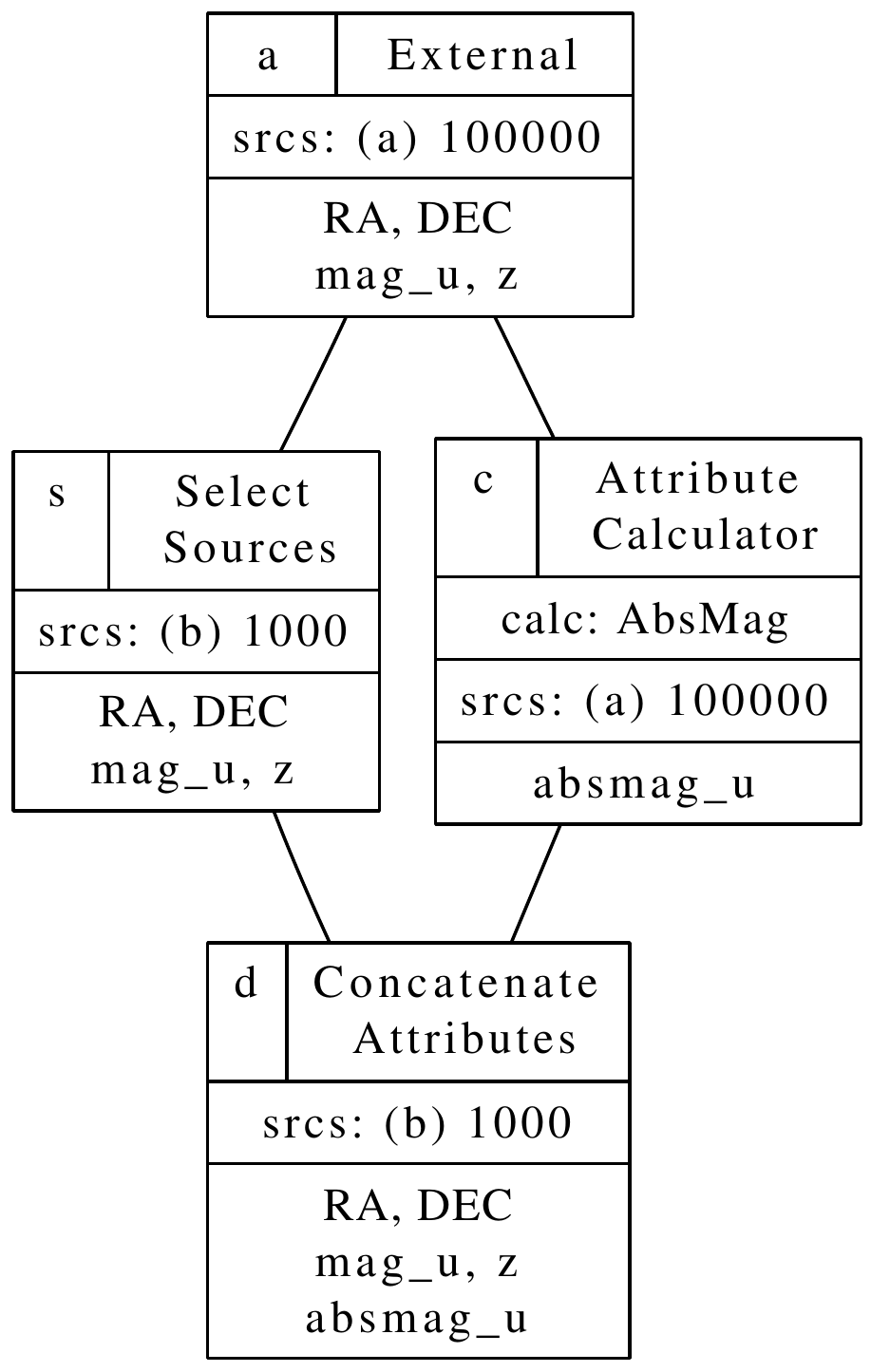}}
\hspace{0.1cm}
 \subfloat[]{\label{fig:scintreexampledata2}\includegraphics[width=\widthscexampleDb]{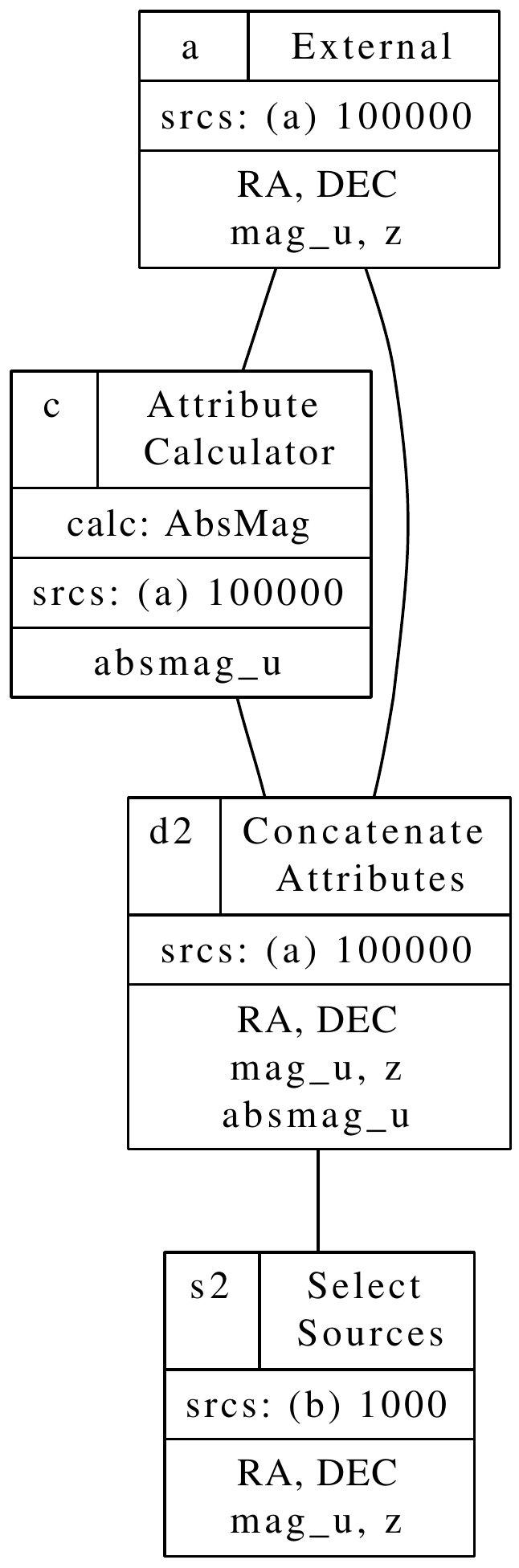}}
\hspace{0.1cm}
 \subfloat[]{\label{fig:scintreexampledata3}\includegraphics[width=\widthscexampleDc]{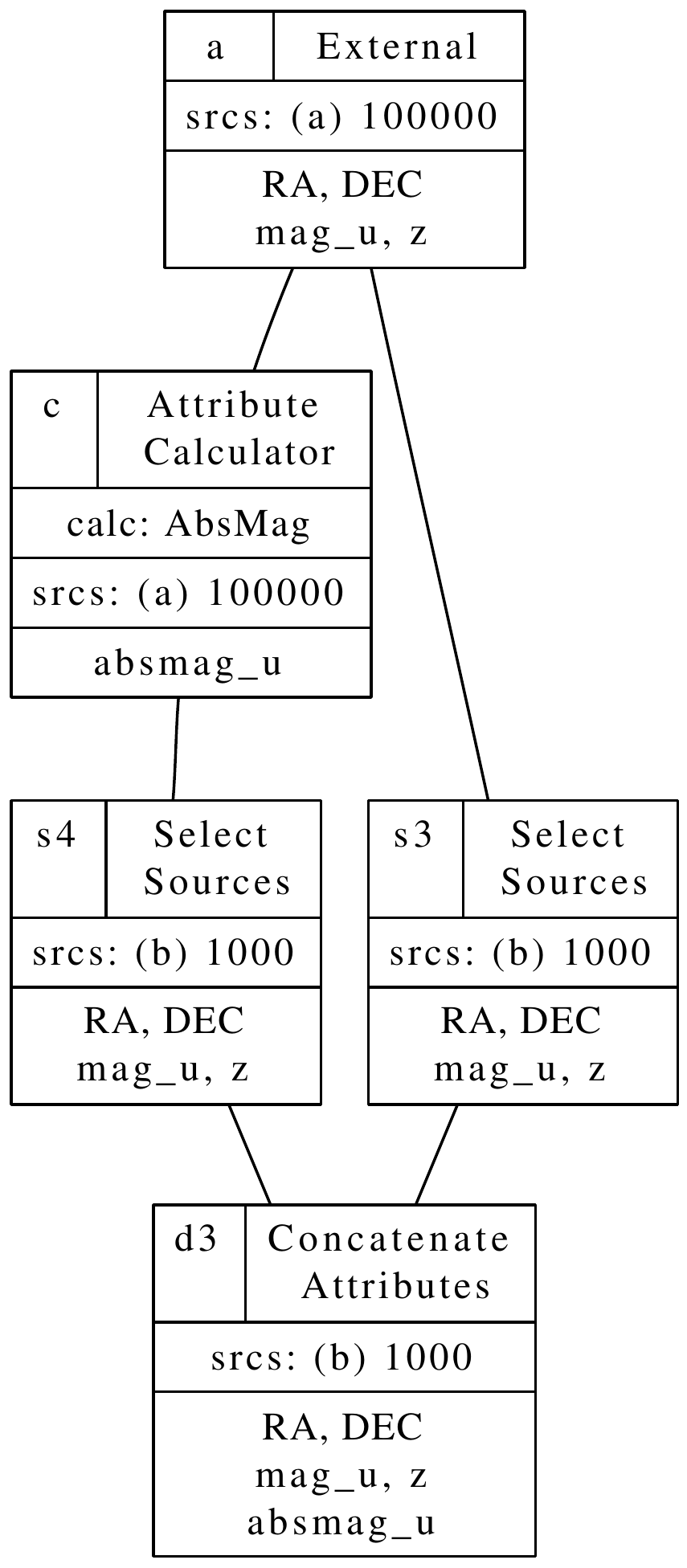}}
\hspace{0.1cm}
 \subfloat[]{\label{fig:scintreexampledata4}\includegraphics[width=\widthscexampleDd]{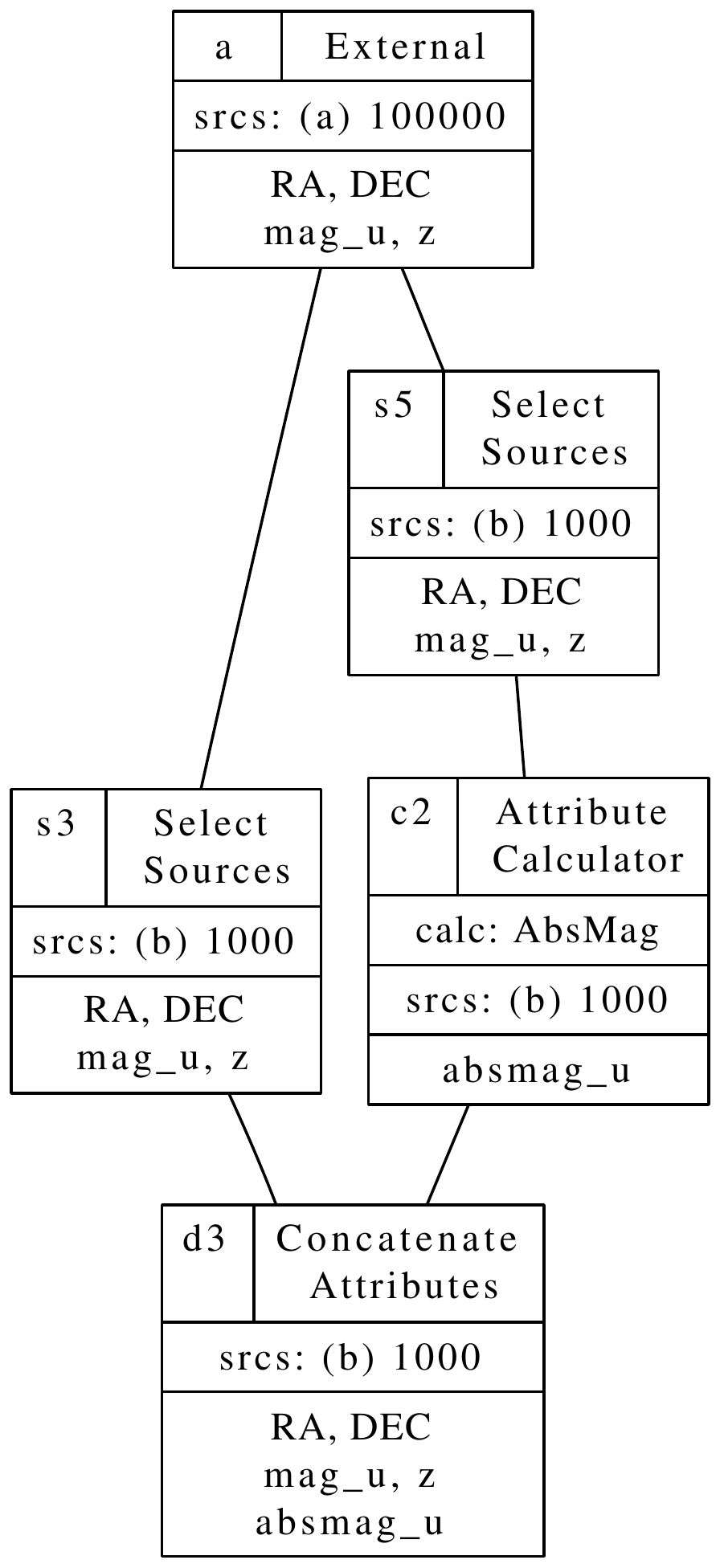}}
\hspace{0.1cm}
 \subfloat[]{\label{fig:scintreexampledata5}\includegraphics[width=\widthscexampleDe]{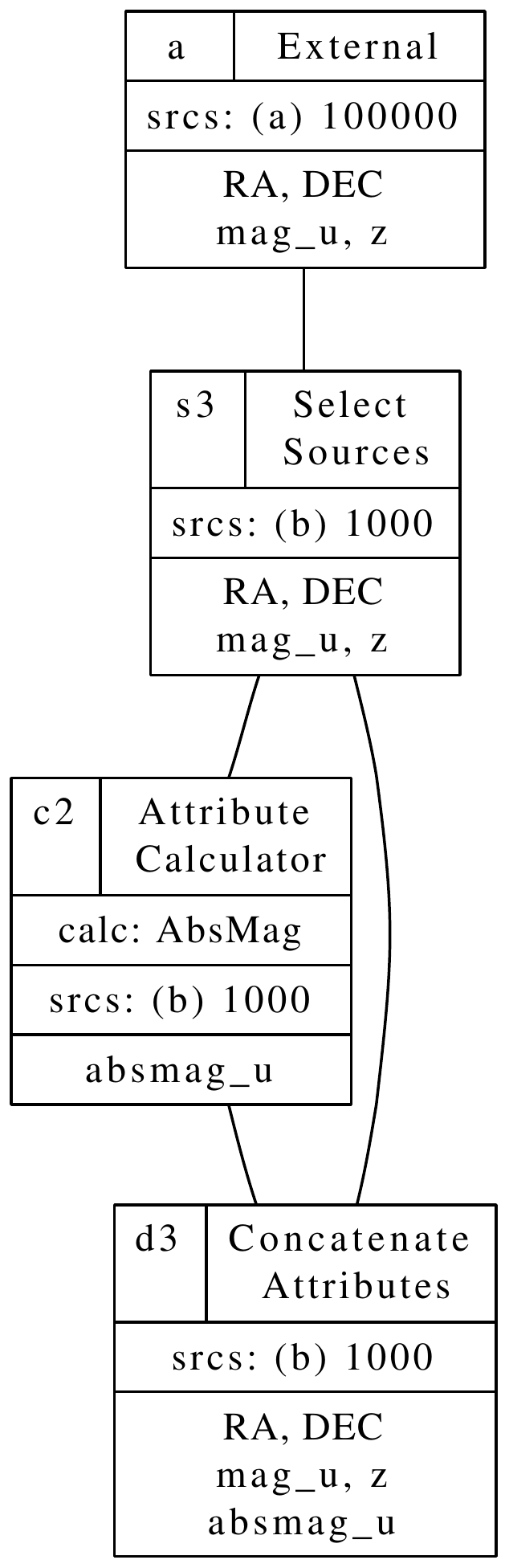}}
\caption{
Source limiting optimizations for before processing. 
On the left a transient copy of the \SourceCollections of \reffigp{fig:scintreexamplepers}, without the final \SelectAttributes. 
The criterion of the \FilterSources has been evaluated so its sources are known and  is converted in a \SelectSources, which explicitly lists the sources.
The \SelectSources can move through the dependency graph freely, since its operator is not dependent on the attributes of the parent anymore.
The \SelectSources is moved down to the end of the dependency graph in figure (b) and subsequently back up in figure (c) and (d). 
In figure (c), two copies of the \SelectSources are created, which are merged back into one in figure (d).
The temporary transient copy \calculator\ $c2$ describes a subset of the original \SourceCollection $C$, thereby reducing the required processing.
}
\label{fig:scintreexampledata}
\end{figure}
 
\end{landscape}

\newcommand{\widthscexampleSa}{0.25\linewidth}
\newcommand{\widthscexampleSb}{0.19\linewidth}
\newcommand{\widthscexampleSc}{0.10\linewidth}
\newcommand{\widthscexampleSd}{0.09\linewidth}

\begin{landscape}
\begin{figure}[ht!]
\centering
 \subfloat[]{\label{fig:scintreexamplesources1}\includegraphics[width=\widthscexampleSa]{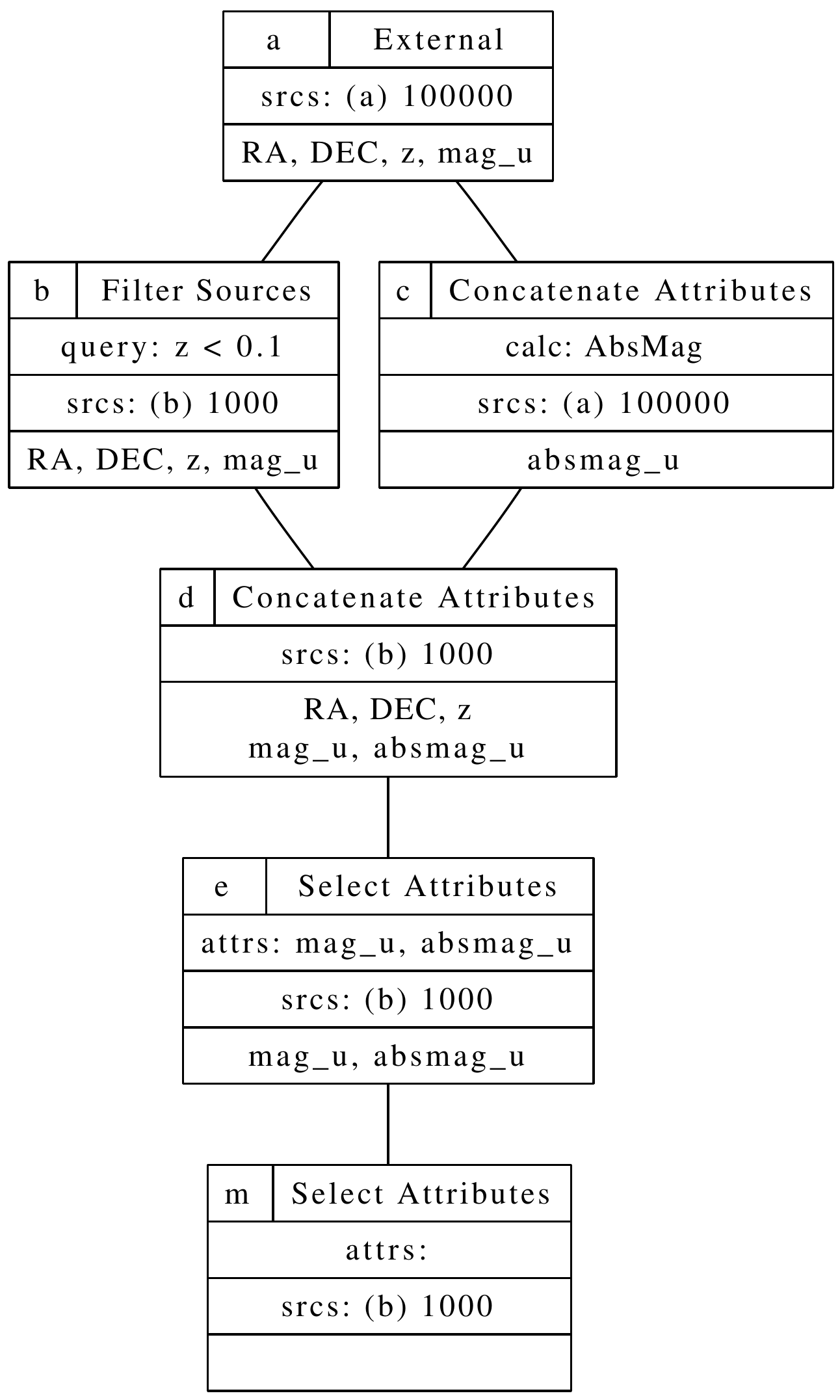}}
\hspace{0.1cm}
 \subfloat[]{\label{fig:scintreexamplesources2}\includegraphics[width=\widthscexampleSb]{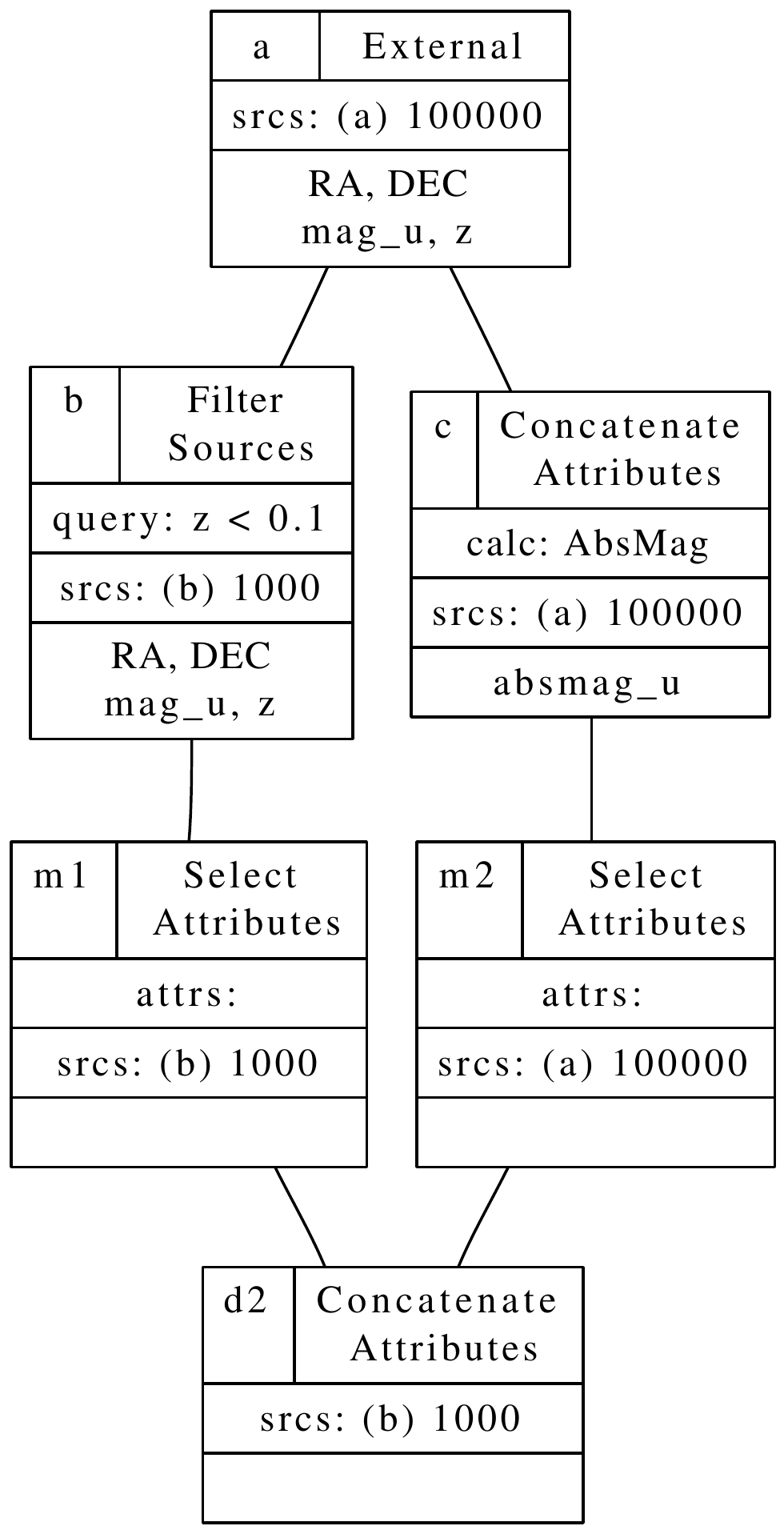}}
\hspace{0.1cm}
 \subfloat[]{\label{fig:scintreexamplesources3}\includegraphics[width=\widthscexampleSc]{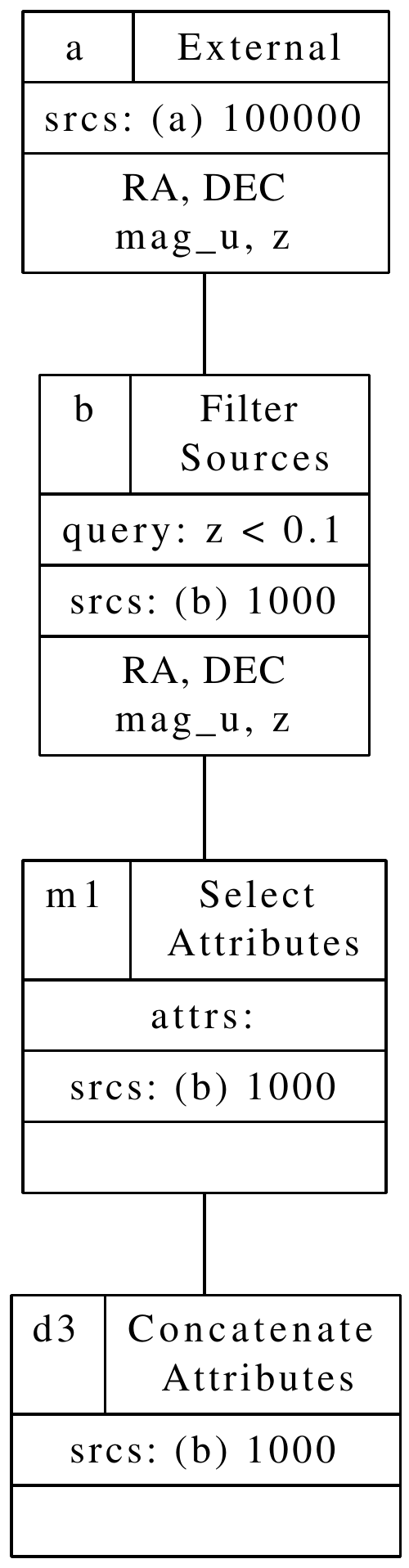}}
\hspace{0.1cm}
 \subfloat[]{\label{fig:scintreexamplesources4}\includegraphics[width=\widthscexampleSd]{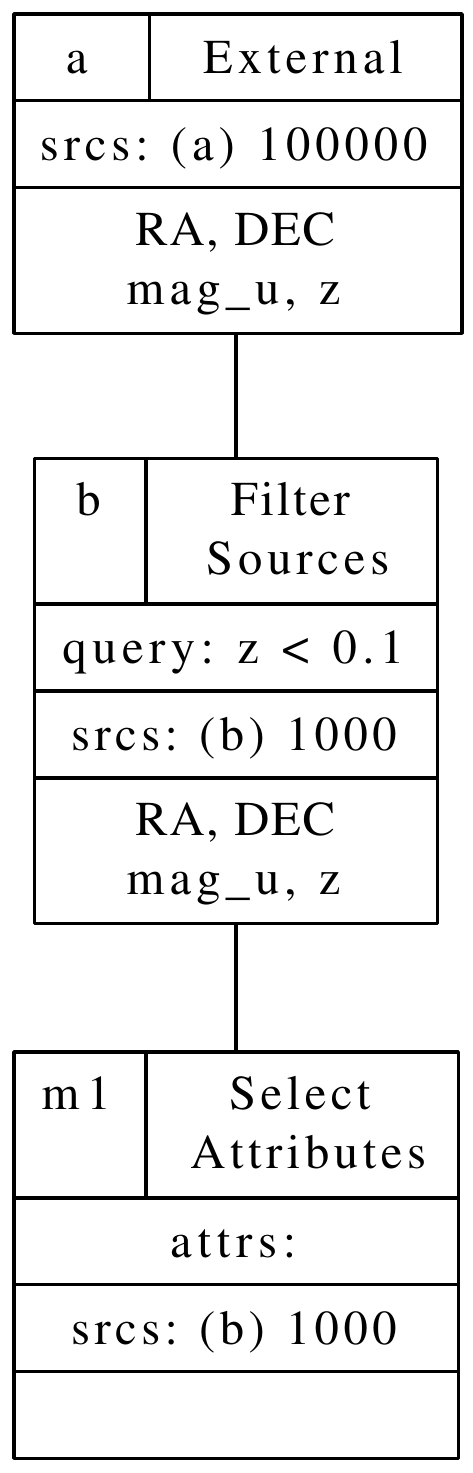}}
\caption{
Evaluation of the exact set of sources in a \SourceCollection.
On the left a transient copy of the \SourceCollections of the example in \reffigp{fig:scintreexamplepers}.
A \SelectAttributes that selects no attributes is placed at the end of the dependency graph, which is subsequently moved up in figure (b).
One of the parents of the \ConcatenateAttributes can be removed, because the algorithm from \citet{setrelationspaper} is used to infer that it is not required anymore.
The \ConcatenateAttributes itself is removed because it has only one parent.
The final dependency graph does not involve any calculation and can be evaluated quickly.
}
\label{fig:scintreexamplesources}
\end{figure}
 
\end{landscape}

\end{document}